\documentclass[manuscript]{acmart}
\AtBeginDocument{%
  }

\usepackage{graphicx}    
\usepackage{caption}     
\usepackage{array} 
\usepackage{tabularx}
\renewcommand{\arraystretch}{1.25} 
\begin{document}

\title{Designing for Dignity while Driving: Interaction Needs of Blind and Low-Vision Passengers in Fully Automated Vehicles}
\author{Zhengtao Ma}
\email{zhengtao.ma@hdr.qut.edu.au}
\orcid{0009-0009-2775-4412}
\affiliation{%
  \institution{School of Design, Queensland University of Technology}
  \city{Brisbane}
  \country{Australia}}

\author{Rafael Gomez}
\email{r.gomez@qut.edu.au}
\orcid{0000-0003-3008-9627}
\affiliation{%
  \institution{School of Design, Queensland University of Technology}
  \city{Brisbane}
  \country{Australia}}

\author{Togtokhtur Batbold}
\orcid{0000-0002-0709-8819}
\affiliation{%
  \institution{School of Computer Science, Queensland University of Technology}
  \city{Brisbane}
  \country{Australia}}
\email{togtokhtur.batbold@hdr.qut.edu.au}

\author{Zishuo Zhu}
\orcid{}
\email{zishuo.zhu@hdr.qut.edu.au}
\affiliation{%
  \institution{Centre for Accident Research and Road Safety - Queensland (CARRS-Q), Queensland University of Technology}
  \city{Brisbane}
  \country{Australia}}

\author{Yueteng Yu}
\email{yueteng.yu@hdr.qut.edu.au}
\orcid{}
\affiliation{%
  \institution{Centre for Accident Research and Road Safety - Queensland (CARRS-Q), Queensland University of Technology}
  \city{Brisbane}
  \country{Australia}}

\author{Ronald Schroeter}
\email{r.schroeter@qut.edu.au}
\orcid{0000-0001-7990-1474}
\affiliation{%
  \institution{Center for Future Mobility, Queensland University of Technology}
  \city{Brisbane}
  \country{Australia}}

\renewcommand{\shortauthors}{Ma et al.}

\begin{abstract}
Fully automated vehicles (FAVs) hold promise for enhancing the mobility of blind and low-vision (BLV) individuals. To understand the situated interaction needs of BLV passengers, we conducted six on-road, and in-lab focus groups with 16 participants, immersing them in real-world driving conditions. Our thematic analysis reveals that BLV participants express a high initial “faith” in FAVs, but require layered, value-sensitive information during the ride to cultivate trust. The participants' modality preference for voice suggests re-evaluating the role of haptics for BLV users in FAVs. Our findings show the importance of a respectful interaction design in FAVs that both address BLV users' mobility challenges and uphold their dignity. While others have advocated for a dignity lens, our contribution lies in grounding this framework in empirical findings and unpacking what it means to design for dignity in the context of FAVs.


\end{abstract}


\begin{teaserfigure}
  \includegraphics[width=\textwidth]{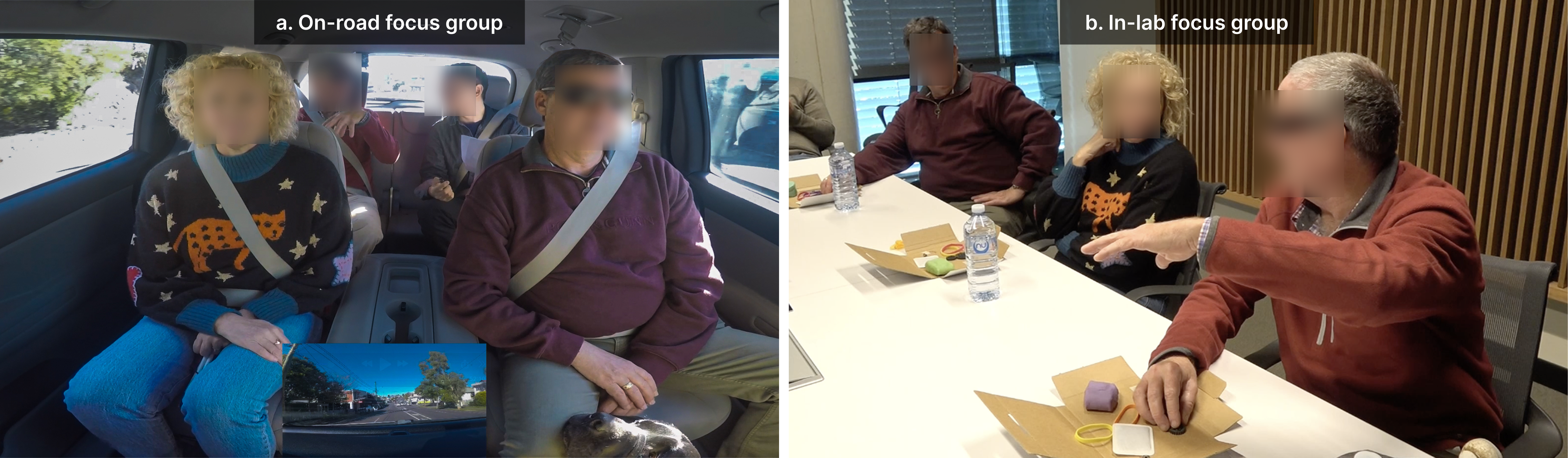}
  \caption{A session with 3 BLV participants in: (a) On-road focus group during a simulated AV ride; (b) In-lab focus group using haptic probes}
  \Description{The figure presents two photographs comparing different focus group settings. Panel (a) shows an on-road focus group, with participants seated inside a moving vehicle and engaging in discussion while an inset road view is visible in the lower corner. Panel (b) depicts an in-lab focus group, where participants are seated around a table in a meeting room, conversing with each other. Both panels illustrate the contrast between conducting discussions in a real driving environment versus an indoor setting.}
  \label{fig:teaser}
\end{teaserfigure}

\received{20 February 2007}
\received[revised]{12 March 2009}
\received[accepted]{5 June 2009}

\maketitle

\section{Introduction}
Fully automated vehicles (FAVs) are promoted as a future mobility solution for users who cannot drive \cite{sae_international_identifying_2019, kassens-noor_autonomous_2021}, including approximately 289 million people worldwide who are blind or have low vision (BLV) \cite{burton_lancet_2021}. As FAVs and highly automated vehicles (HAVs, such as Waymo Robotaxi \footnote{Waymo is a self-driving ride-hailing service that uses autonomously driven vehicles, without a human driver.}) advance toward minimal human intervention \cite{J3016_202104}, the prospect of improved mobility independence for BLV communities is becoming increasingly realistic. 

Prior research has shown BLV users' willingness to adopt FAVs and their recognition of the potential benefits for travel independence \cite{bennett_willingness_2020}. Several surveys and focus groups have further examined acceptance and concerns among BLV users \cite{brinkley_opinions_2017, brinkley_exploring_2020},  while researchers have highlighted key information needs and proposed human-machine interface (HMI) designs, such as support for safety while in ride \cite{fink_expanded_2023, fink_autonomous_2023} and boarding \cite{Meinhardt_lightmyway_2025}. 

However, since FAVs have not yet been deployed in real-world settings, most findings stem from hypothetical or vision-dominant simulation studies.  Traditional driving simulators, typically designed for sighted drivers, fail to reproduce critical non-visual sensory modalities (e.g., sound, haptics, and body motion) that strongly shape BLV passengers' sense of safety and independence. Scholars, therefore, highlight the importance of situated studies that account for real driving conditions and strengthen ecological validity \cite{baltodano_rrads_2015, meurer_wizard_2020}.

In addition, non-visual interaction modalities in FAVs remain underexplored \cite{jansen_design_2022}. While auditory, mainly voice, has proven effective \cite{brinkley_open_2019}, \textit{haptics}, which has long been promoted as a promising channel for BLV accessibility \cite{sucu_haptic_2013, jansen_design_2022}, has been surrounded by both promise and scepticism, with limited evidence of its standalone benefits and added-on value when combined with other modalities in FAV contexts. Questions remain about what information haptics should convey and how effective such interactions are in practice from the perspective of BLV users. 

To address these gaps, we conducted six focus groups with 16 BLV participants, combining on-road and in-lab sessions, and used a human-driven people-mover to simulate the experience of driving in a FAV. This setup exposed participants to authentic vehicle dynamics and environmental cues. To further explore non-visual modalities, we introduced physical haptic probes that represent the six modalities in detail: (1) vibrotactile; (2) surface friction and skin stretch; (3) thermal tactile; (4) mid-air haptics; (5) shape-changing or deformable interfaces; and (6) force feedback, to stimulate embodied reflection right after the driving.

We investigate:
\begin{itemize}
    \item \textbf{RQ1:} What are the pragmatic and hedonic needs of BLV passengers during FAV rides?
    \item \textbf{RQ2:} How can non-visual interactions, particularly haptics, support these needs and shape user experience?
\end{itemize}

Our findings identify BLV users' initial "faith" in FAVs, and reveal that BLV passengers' needs extend beyond pragmatic safety to include connection, control, and the hedonic dimensions of social participation and the psychological pleasure of “driving.” We show how voice interaction builds on existing competencies, while the assumed benefits of immediate haptic cues often fail to materialize without clear interpretability and actionability. By situating these insights in real-world conditions, this study offers empirical evidence about BLV users' trust, haptic value, and interaction needs in FAVs. We argue for a dignity lens framework that reframes accessible FAVs for BLV users not only as providers of functional mobility, but also as vehicles for affirming autonomy, minimizing cognitive intrusiveness, respecting competence, enabling emotional and social fulfillment, and supporting healthy interdependence. 

\textit{Contribution statement}: We contribute empirical insights through qualitative evidence on how BLV passengers experience FAV travel, with a nuanced understanding of their trust dynamics, layered interaction needs, and interaction modality preferences in situated contexts. We foreground a \textit{Designing for Dignity} perspective in the FAV context, offering practical suggestions for dignity-oriented interactions in FAVs.  

\section{Related work}
\paragraph{Technology background: FAVs and HAVs}
According to SAE standards \cite{J3016_202104}, FAVs (Level 5 automation) do not require any human driving maneuvers and are capable of self-driving under all conditions. In contrast, HAVs (Level 4 automation) can operate autonomously only within certain conditions and defined operational domains, often bounded by a digital fence. Both types of AVs may be deployed in different ownership models, ranging from shared ride-hailing services, in the form of robotaxis, to privately owned concepts such as Tesla's proposed approach. In this study, we broadly discuss FAVs of all types of ownership.

\subsection{BLV users in FAVs}
As a newly emerging user group in the context of AVs, BLV communities have expressed strong interest and willingness to use FAVs \cite{brewer_supporting_nodate}. Bennet et al.'s \cite{bennett_willingness_2020} survey with 211 BLV individuals found that willingness to adopt FAVs is strongly motivated by the hope of independence and travel freedom, but also tempered by concerns over safety and affordability. Although such quantitative surveys \cite{brinkley_exploring_2020, fink_give_2023} give us a valuable view of the landscape in terms of BLV attitudes and concerns, participatory studies help us understand their interaction needs and personal/social factors.

Fink et al. report through focus groups that information about the route and surrounding environment helps BLV participants feel connected during travel \cite{fink_expanded_2023}. Moreover, Meinhardt et al. show that information needs extend from strategic routes to traffic information \cite{meinhardt_hey_2024}, to more safety-related exit obstacles \cite{Meinhardt_lightmyway_2025}. While previous studies have considered information needs, Brewer et al. highlight that such factors as the sense of control are desired by BLV passengers \cite{brewer_understanding_2018}.

However, prior research on BLV travelers' information needs has largely been based in laboratory settings. While effective for producing comprehensive lists of BLV users' needs, these approaches offer limited ecological validity and tend to elicit abstract preferences rather than capturing situated user responses. As a result, a knowledge gap still exists in a deeper and more nuanced understanding of the interaction needs of BLV users in a real-world driving context.

\subsection{Non-visual HMI for BLV users}
HMI design for BLV riders' accessibility in future FAVs has converged on a common objective: leveraging non-visual modalities to meet BLV users' information needs and ensure safety during the entire use \cite{brewer_supporting_nodate,huff_participatory_2020,fink_fully_2021}. Early explorations on HMIs leveraged haptic, auditory, and multimodal interactions.

\subsubsection*{Haptic HMI}
While haptics are widely adopted in assistive technologies for vision impairment \cite{hapticAT_Sorgini19052018}, only a small but growing body of work has explored its potential in the broader context of AVs. Explorations, such as \citet{sucu_haptic_2013}'s vibrotactile steering wheel, provided turn-by-turn navigation for BLV users on a simulated drive. Although manual driving is unnecessary in FAVs, these navigation concepts have carried forward into their impact on BLV passengers' user experiences.

Motivated by user needs, current haptic HMI research for BLV users has largely focused on two challenges: (1) vehicle finding, onboarding and offboarding \cite{fink_autonomous_2023, Meinhardt_lightmyway_2025}, and (2) situational awareness during the ride \cite{ranjbar_vibrotactile_2022, meinhardt_hey_2024}. Ranjar et al. developed vibrotactile pre-cues to enhance situational awareness for BLV passengers, and found that significant cognitive effort was required to memorize vibration patterns \cite{ranjbar_vibrotactile_2022}. Building on participatory studies of BLV user needs, Fink et al. explored ultrasonic mid-air haptics to present touchable intersection shapes \cite{fink_expanded_2023, Fink_Autonomous_2023_midair}, offering insights into combining natural gesture interaction with haptic feedback. More recent studies extend haptic interaction beyond isolated features to encompass the complete ride accessibility: their prototypes use physical and shape-changing forms, with OnBoard \cite{meinhardt_hey_2024} presenting vehicle behavior, traffic, and route information; and PathFinder \cite{Meinhardt_lightmyway_2025} providing exit and obstacle cues. 



\subsubsection*{Auditory HMI}
A vast literature exists in the context of auditory accessibility interfaces for BLV users for a wide range of everyday contexts \cite{Accessibility_Kelly_2021}, so naturally, FAV research has explored HMIs that enable voice-based interaction for BLV users as well. For instance, Brinkley et al. introduced ATLAS \cite{brinkley_open_2019}, a system that supports situational awareness by providing audible location cues during transit in FAVs. An on-road evaluation with BLV participants demonstrated that this interface supported situational awareness and optimized the overall usability of FAVs. 

\subsubsection*{Multimodal HMI}
Many studies, including those mentioned above, combine haptics with auditory channels to support an effective interaction with AVs. For example, the AVA system \cite{fink_autonomous_2023} used vibration and sound feedback to help users locate the vehicle, while other haptic artifacts have been paired with voice interaction to provide a language-based input channel and convey semantic information \cite{Fink_Autonomous_2023_midair, meinhardt_hey_2024}. Yet, evidence remains mixed: for example, Fink et al. found that the "+haptic" interfaces did not yield significant improvements over voice-only designs in a navigation task \cite{Fink_Autonomous_2023_midair}, while Meinhardt et al. also reported a higher mental workload while interacting with OnBoard \cite{meinhardt_hey_2024}. 

Although multimodal HMIs are increasingly being studied, comparatively, the role of haptics in FAV accessibility for BLV passengers is still underexplored. Whether the “sense of touch” can supplement the kinds of information that sighted passengers obtain visually, such as spatial layouts, maps, or environmental cues, remains an open question and a persistent myth in the discourse on non-visual interaction.

\subsection{Qualitative on-road \& in-vehicle studies}
Automotive researchers have long encouraged on-road studies for their ability to provide a more realistic setting compared to simulation \cite{Dirk_Ghost_2015, meurer_wizard_2020, Faramarzian_Anthropological_2021}. For BLV participants in particular, the limitations of driving simulators are especially evident: most simulators are vision-dominated and fail to capture the realism of sound, motion, smell, light, traffic, and urban texture \cite{CARSTEN201187}. 
To balance ecological validity with safety, researchers have developed a range of Wizard-of-Oz (WoZ) methods for on-road FAV studies. The RRADS system \cite{baltodano_rrads_2015} was among the first to explore autonomous driver–vehicle interaction using commercially available cars on public roads. Building on RRADS, Marionette WoZ introduced false controllers, that allowed participants to interact with a non-functional steering wheel \cite{wang_marionette_2017}. Similarly, Detjen et al. proposed a WoZ setup in which a large screen in a people-mover concealed the driver to study non-driving-related tasks in FAVs \cite{detjen_wizard_2020}. 

Although in-situ on-road setups introduce a degree of unpredictability, this lack of control is often valued in highly qualitative research on the situated user experience and activities. It allows researchers to observe how users actually feel and behave when exposed to everyday driving conditions, revealing underlying frictions and lived experiences that go beyond evaluating the static usability of specific interfaces.


\section{Method}
Given the exploratory nature and naturalistic set-up of this study, we employed focus groups to elicit diverse perspectives and foster a sense of “community” among BLV participants. The study comprised two sessions: (1) an on-road focus group, addressing RQ1, conducted in a simulated FAV operated by a human driver to create an in-situ environment of travel; and (2) an in-lab focus group, addressing RQ2, which incorporated haptic probes to evoke a sense of haptic interaction.

\subsection{Participants}
Prior to participant recruitment, we received human research ethics approval from the university (approval \#9361). Participant recruitment was facilitated through two primary channels: (1) a recruitment post on a public BLV community group, and (2) promotion by local and national BLV organizations via email and newsletters. Recruitment materials, including posts and the expression of interest forms (Qualtrics), were reviewed by disability advocates to ensure digital accessibility. 

Participants were eligible if they were aged 18 years or older and self-identified as blind or low vision. To engage the target user group and safeguard participant well-being, we excluded individuals who currently held a valid driver's license, as our focus was on those who had lost the freedom to drive (acknowledging that in some jurisdictions, people with certain levels of low vision may obtain conditional licenses). We also excluded individuals with pre-existing health conditions, such as chronic motion sickness or vestibular disorders, that could adversely affect their comfort during vehicle travel. 

Also, participants were welcome to attend with support workers or guide dogs, both to ensure safe access to the study site and to foster a more familiar and comfortable group atmosphere. Each participant received a physical grocery gift card (valued at \$100) as an incentive in recognition of their contribution. 

In the end, 16 BLV participants (N = 16, 5 female, 11 male) with an average age of 56 years participated and were divided into six groups (maximum 3 participants in one group). Shown in Table \ref{tab:participants}, seven participants were totally blind, while the remainder reported diverse low-vision conditions. None were newly vision-impaired; six were congenitally blind or with low vision, ten experienced progressive vision loss later in life, and one brought their guide dog. Regarding driving experience, seven participants had legally driven prior to vision loss, and two participants (though never licensed drivers) reported experience with learner driving or operating golf carts on test tracks. 

\begin{table}[htbp]
\centering
\caption{Participant demographics and vision impairments}
\label{tab:participants}
\footnotesize
\begin{tabularx}{\linewidth}{|c|c|c|c|c|p{3.5cm}|p{3.5cm}|c|}
\hline
\textbf{G ID} & \textbf{P ID} & \textbf{Age} & \textbf{Gender} & \textbf{Vision code} &
\textbf{Self-reported vision impairment} & \textbf{Major trans. use} &
\textbf{Road driving} \\
\hline
1 & 1  & 35 & M & Born blind       & Born blind & Driven by parent; Buses; Trains & No(Test track only)\\
\hline
1 & 2  & 59 & F & Blind            & Totally blind, lost sight later & Driven by partner & Yes \\
\hline
2 & 3  & 63 & M & Low vision       & Very low vision now, lost sight later & Buses & Yes \\
\hline
2 & 4  & 54 & F & Low vision       & Retinitis pigmentosa (RP), $\sim$7–8$^\circ$ central vision, poor peripheral & Buses; Trains & Not disclosed \\
\hline
2 & 5  & 53 & M & Legal blind      & Chromatopia, no daytime vision, no colour perception & Driven by support workers; Buses & Yes (also test track)\\
\hline
3 & 6  & 61 & M & Born low vision  & Born with RP, fully blind for 20 years & Driven by partner; Buses & No \\
\hline
3 & 7  & 60 & M & Low vision       & RP later in life, near end-stage & Driven by support workers and child; Buses & Yes (also test track)\\
\hline
4 & 8  & 50 & M & Born low vision  & Since birth, tunnel vision ($<$5\%) & Driven by support workers; Buses & No(Test track only)\\
\hline
4 & 9  & 32 & M & Born blind       & Glaucoma + Peter's anomaly & Cabs/Uber; Driven by support workers; Buses & No \\
\hline
4 & 10 & 77 & M & Blind            & Totally blind for 16 years & Driven by partner & Yes \\
\hline
5 & 11 & 53 & M & Born blind       & Premature birth, totally blind & Buses; Driven by partner & No \\
\hline
5 & 12 & 60 & F & Born low vision  & Born with low vision & Buses; Driven by child & No \\
\hline
5 & 13 & 49 & F & Low vision       & RP diagnosed at age 25, now 49 & Driven by partner; Buses & No \\
\hline
6 & 14 & 63 & F & Low vision       & Diabetic retinopathy, 3$^\circ$ vision left eye & Driven by partner & Yes \\
\hline
6 & 15 & 58 & M & Low vision       & RP diagnosed at age 26 & Driven by partner; Buses & Yes \\
\hline
6 & 16 & 70 & M & Born blind       & Born partially blind, lost sight at 14 & Driven by support worker & No \\
\hline
\end{tabularx}
\end{table}

\subsection{Apparatus}
\subsubsection{On-road focus group}
\begin{figure}
    \centering
    \includegraphics[width=0.5\linewidth]{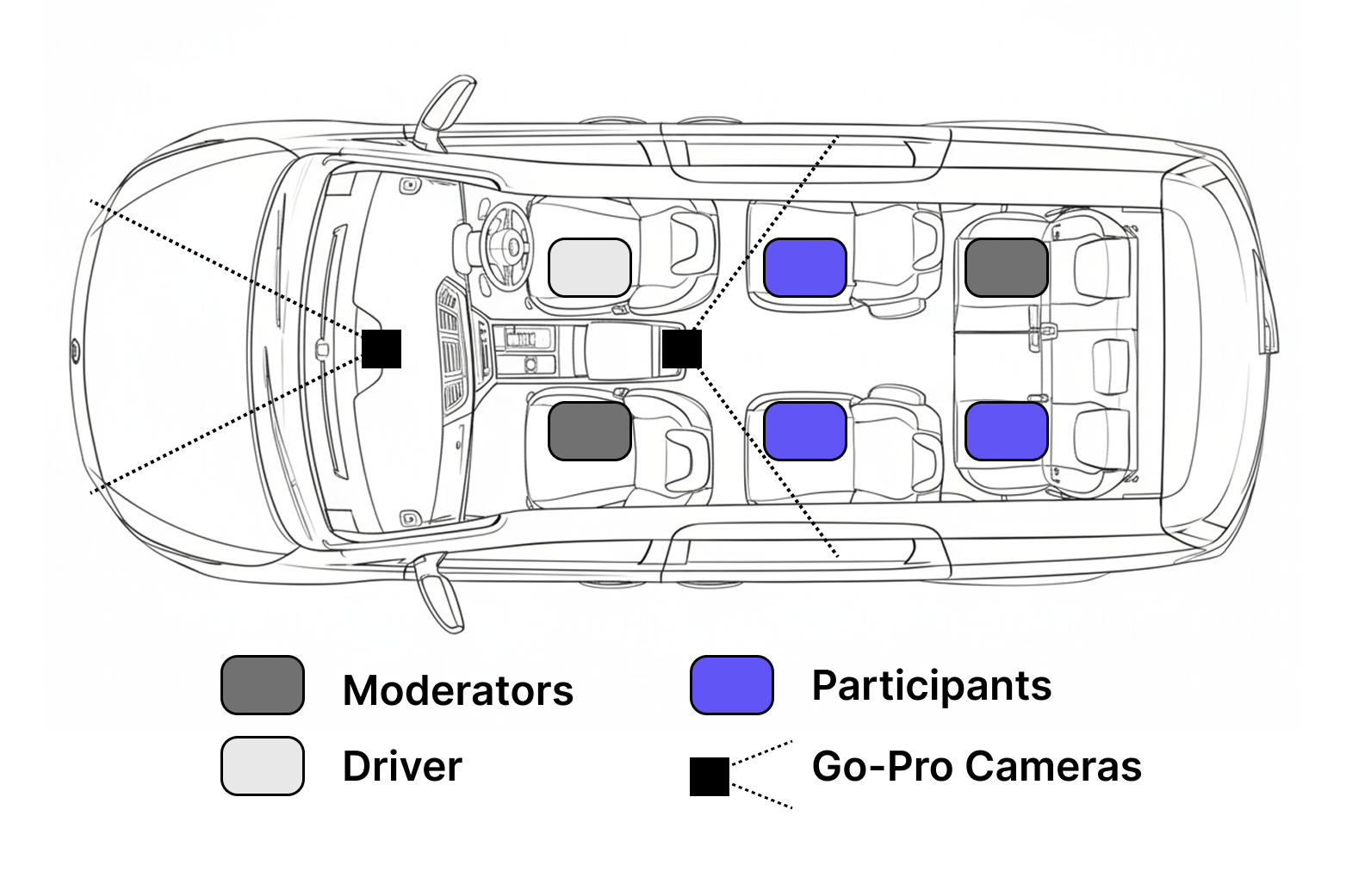}
    \caption{Schematic of people mover for the on-road focus group}
    \Description{Schematic of the eight-seat people mover used on the road. It marks the driver, the assistant in the front passenger seat, the moderator in the third row, and two participants in the second row, one participant in the third row. It also shows two GoPro cameras: one at the front windshield towards the road, another at the middle armrest of the front seat towards the cockpit. The setup supports discussion during motion.}
    \label{fig: car set-up}
\end{figure}
The study was conducted in an eight-seat People mover (Kia Carnival), operated by an experienced sighted driver. This vehicle was selected to provide ample interior space, support participant comfort, and enable discussions to unfold naturally during the ride. Two participants were seated in the second row and one in the third row. The main moderator facilitated the discussion from the third row, while an assistant moderator in the front passenger seat provided logistical support throughout the session. The vehicle followed a predefined route at legal speed limits, with the driver serving as a “driving wizard” to simulate the behavior of a FAV. To preserve the illusion of automation, no interaction occurred between the driver and participants. A schematic of the vehicle setup is shown in Fig.~\ref{fig: car set-up}.

Two GoPro Action 2 cameras with audio recording capabilities were installed inside the vehicle. One “road-cam” was mounted on the front windshield to capture the external driving environment, while a second “cockpit-cam” was mounted behind the front passenger seat to record participants' verbal discussions and physical behaviors. To ensure high-quality audio despite road noise, a backup microphone was also affixed to the seating area.

\subsubsection{In-lab Focus Group}
For the in-lab data collection, a stationary video recorder was positioned in the study room to capture the pre- and post-ride focus group sessions (see Fig.~\ref{fig:teaser}).

\begin{figure}
    \centering
    \includegraphics[width=0.75\linewidth]{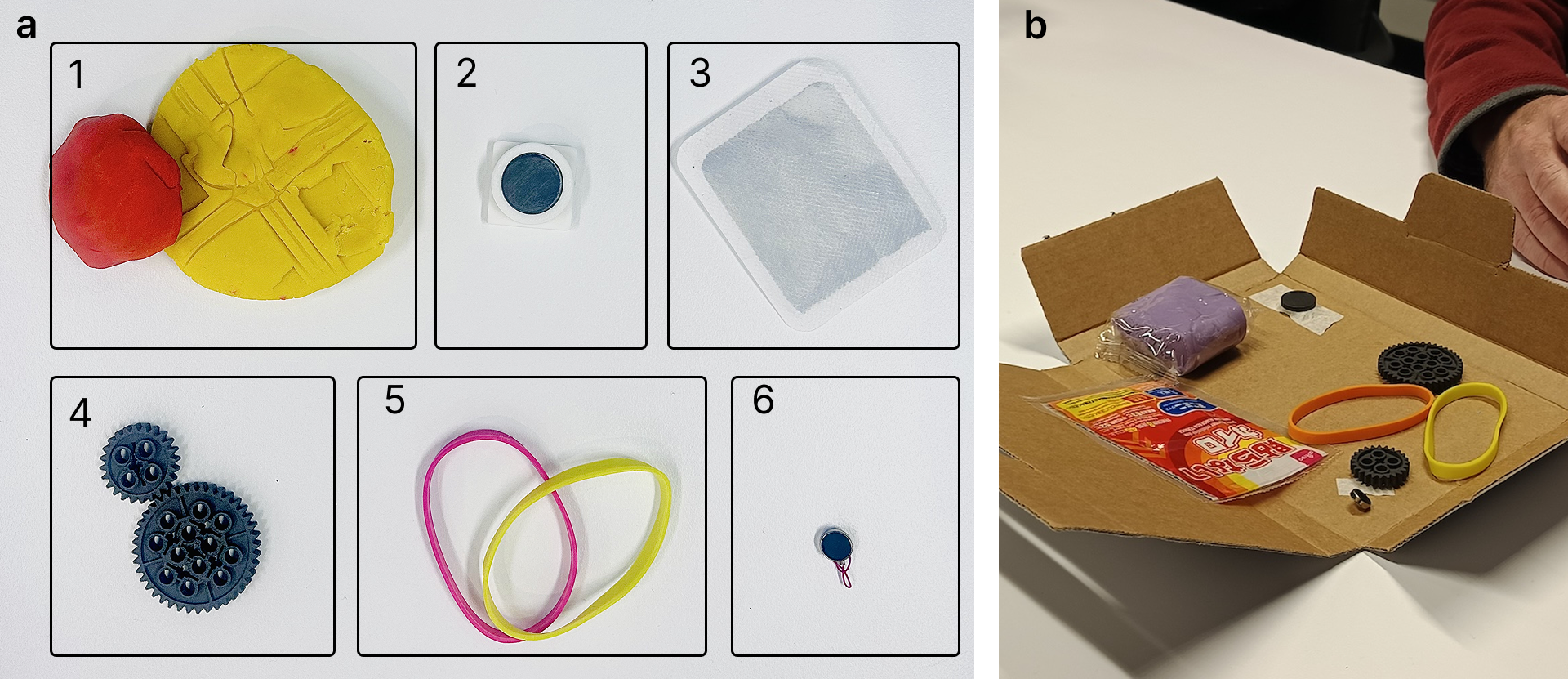}
    \caption{Haptic probes based on the six haptic interaction modalities. a) Schematic view of probes: (1) shape-changing or deformable interfaces, (2) mid-air haptics, (3) thermal tactile, (4) surface friction and skin stretch, (5) force feedback, and (6) vibrotactile. b) the probes given to participants. }
    \Description{Two-panel image showing haptic probe materials used in a design study. Panel (a) shows six representative probes on the table: Play-Doh modeling clay, Magnet pair, Heat pad, Gears, Rubber band, and Coin vibrator. Panel (b) shows the physical probes laid out for participants.}
    \label{fig: haptic probs}
\end{figure}

\textbf{Haptic Probes.} Embodied interaction and sensory engagement are crucial in participatory design with BLV users, as evident in contexts such as body movement \cite{silva_dance_2025} and tangible ideation \cite{meinhardt_hey_2024}. Drawing on these practices, we assembled a set of physical probes to illustrate a wide range of haptic sensations. These probes (in Fig\ref{fig: haptic probs}) were designed to be off-the-shelf and easily understandable, minimizing learning effort and reducing bias toward any particular existing haptic technology. The probe set simulated six distinct haptic modalities commonly referenced in AV interaction research \cite{jansen_design_2022}: (1) vibrotactile, (2) surface friction and skin stretch, (3) thermal tactile, (4) mid-air haptics, (5) shape-changing or deformable interfaces, and (6) force feedback. Table~\ref{tab:haptic_probs} provides a detailed description of these probes, the modalities they represent, and example descriptions provided to participants during the sessions.

\begin{table}[h!]
    \centering
    \caption{Descriptions of various haptic technologies.}
    \label{tab:haptic_probs}
    \renewcommand{\arraystretch}{1.3} 
    \footnotesize
    \begin{tabularx}{\textwidth}{|p{0.22\textwidth}|p{0.22\textwidth}|X|}
        \hline
        \textbf{Haptic Type} & \textbf{Object} & \textbf{Description} \\
        \hline
        Vibrotactile & Coin vibrator & ``Put your hand on this box to feel the vibration.'' \\
        \hline
        Surface Friction \& Skin Stretch & Gears & ``Touching the edge of gears, feel the friction and skin stretch.'' \\
        \hline
        Thermal & Heat pad & ``Touch the package, feel the warmth on its surface.'' \\
        \hline
        Mid-Air Haptics & Magnet pair & ``Hold one magnet in your hand, bring it close to another on the table, feel the tension in the air.'' \\
        \hline
        Shape-Changing / Deformable & Play-Doh & ``Touch this soft clay, imagine the shape changing and responding to you.'' \\
        \hline
        Force Feedback & Rubber band & ``Feel the force of the rubber band resisting you.'' \\
        \hline
    \end{tabularx}
\end{table}

\subsection{Procedure}
Each focus group session lasted approximately 2.5 hours, comprising three sequential components: pre-ride, on-road, and in-lab discussions (see Table~\ref{tab:study1agenda}).

\begin{table}[!htbp]
\renewcommand{\arraystretch}{1.2}
\caption{Scheduled agenda for the two-and-a-half-hour focus group}
\label{tab:study1agenda}
\begin{tabularx}{\linewidth}{|>{\centering\arraybackslash}p{2.2cm}|>{\raggedright\arraybackslash}X|}
\hline
\textbf{Duration} & \textbf{Agenda} \\
\hline
5 min  & Study introduction and consent form signing (in lab). \\
\hline
30 min & Pre-ride: ice-breaker, participant self-introductions, and discussion of transportation needs and attitudes toward FAVs. \\
\hline
5 min  & Transition to the vehicle and familiarization with seating and layout. \\
\hline
50 min & On-road focus group: discussion of information needs and think-aloud reflection during the ride. \\
\hline
50 min & Post-ride focus group (in lab): revisit information needs and explore preferred interaction modalities. \\
\hline
10 min & Session summary and distribution of incentives. \\
\hline
\end{tabularx}
\end{table}

\subsubsection{Pre-ride Focus Group (30 minutes)}
Following the study introduction and consent process, participants were invited to share their perceptions, prior knowledge, and expectations regarding FAVs, addressing RQ1. To prompt reflection, an audio-described promotional video of Waymo\footnote{Waymo video used: \textit{“[Audio Described] LightHouse for the Blind SF CEO Shares Why She Rides with Waymo”}, available at: \url{https://youtu.be/XrBiji8mjDg?si=MeJdEMlOzO5Xb6Rp} (accessed July 2025).} was shown as an illustrative example of state-of-the-art AV applications. The discussion focused on transportation challenges to BLV individuals, including access barriers, reliance on social support, and expectations for independent mobility. This session helped establish contextual grounding for the subsequent on-road experience.

\subsubsection{On-road Focus Group (50 minutes)}
The on-road component was designed as an in-situ elicitation of real-time experiences within a moving vehicle. Across the 50-minute session, participants were encouraged to "think aloud" and comment on their impressions, information needs, and embodied responses during the ride. This phase served as a naturalistic probe, allowing participants to express immediate reactions as well as recall prior experiences triggered by the journey. Discussion topics encompassed both pragmatic concerns (e.g., accessibility, functionality, safety) and hedonic aspects (e.g., trust, comfort, enjoyment), with particular attention to vehicle dynamics, environmental factors, and non-visual interface modalities. The moderators ensured balanced participation and probed for elaboration on emerging topics.

\begin{figure}
    \centering
    \includegraphics[width=0.5\linewidth]{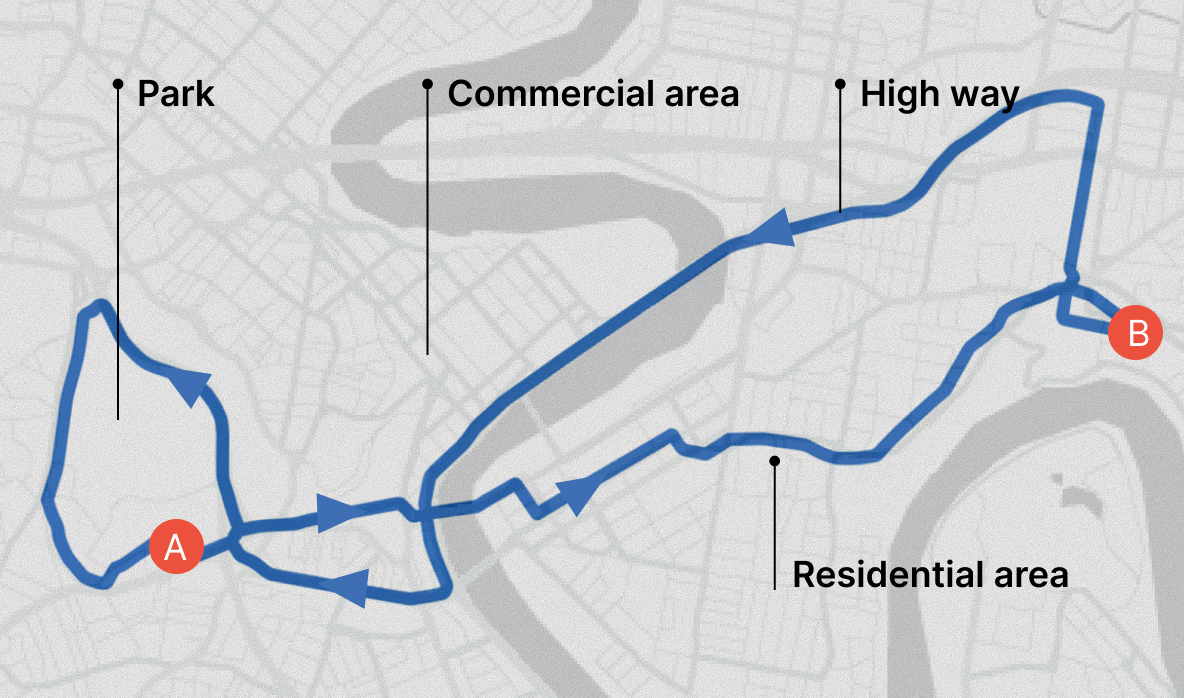}
    \caption{Map of the on-road route and passing areas.}
    \Description{A simplified city map shows a driving route marked in blue, beginning at point A on the left and ending at point B on the right. The route loops through several environments: a park near the starting point, a commercial area in the upper middle, a stretch along a highway toward the northeast, and a residential area in the lower middle before reaching point B. Arrows along the route indicate the direction of travel.}
    \label{fig:route design}
\end{figure}

\paragraph{Route Design} The driving route followed a loop with one midway break to ensure participant comfort and reduce fatigue. It included a mix of urban and suburban roads, civic and natural surroundings, and diverse functional zones (e.g., business and residential districts) to mirror the varied transportation contexts encountered by BLV users. The route was also designed to expose participants to a range of traffic conditions and orientation scenarios, including highways, roundabouts, and multi-turn intersections. The vehicle travelled from the starting Point A to Point B (approx. 20 minutes), pausing briefly for a break (approx. five minutes), and then returned to the starting point via a different path (approx. 25 minutes), shown in Fig\ref{fig:route design}.


\subsubsection{In-lab Focus Group (50 minutes)}
To address RQ3, the final in-lab session focused on interaction preferences (particularly haptic input/feedback) in the context of FAVs. Participants first revisited issues raised during the ride, then shifted to articulating desired modes of interaction. At this stage, the haptic probes were introduced to support embodied exploration. Participants engaged with these materials while expressing feedback both verbally and through touch. This process enabled participants to associate specific haptic modalities with layered information needs.

\subsection{Analysis}
The dataset comprised over 15 hours of audio and video recordings, which consisted of: (1) pre-ride focus groups; (2) on-road discussions; (3) synchronized road-cam and cockpit-cam video; and (4) in-lab sessions. Audio was transcribed with Amazon AI Transcribe (as approved by ethics and the university) and corrected by the authors. Video was synchronized with transcripts and used to contextualize behaviors (e.g., gestures, body movements, tactile engagement).

\begin{figure}
    \centering
    \includegraphics[width=0.85\linewidth]{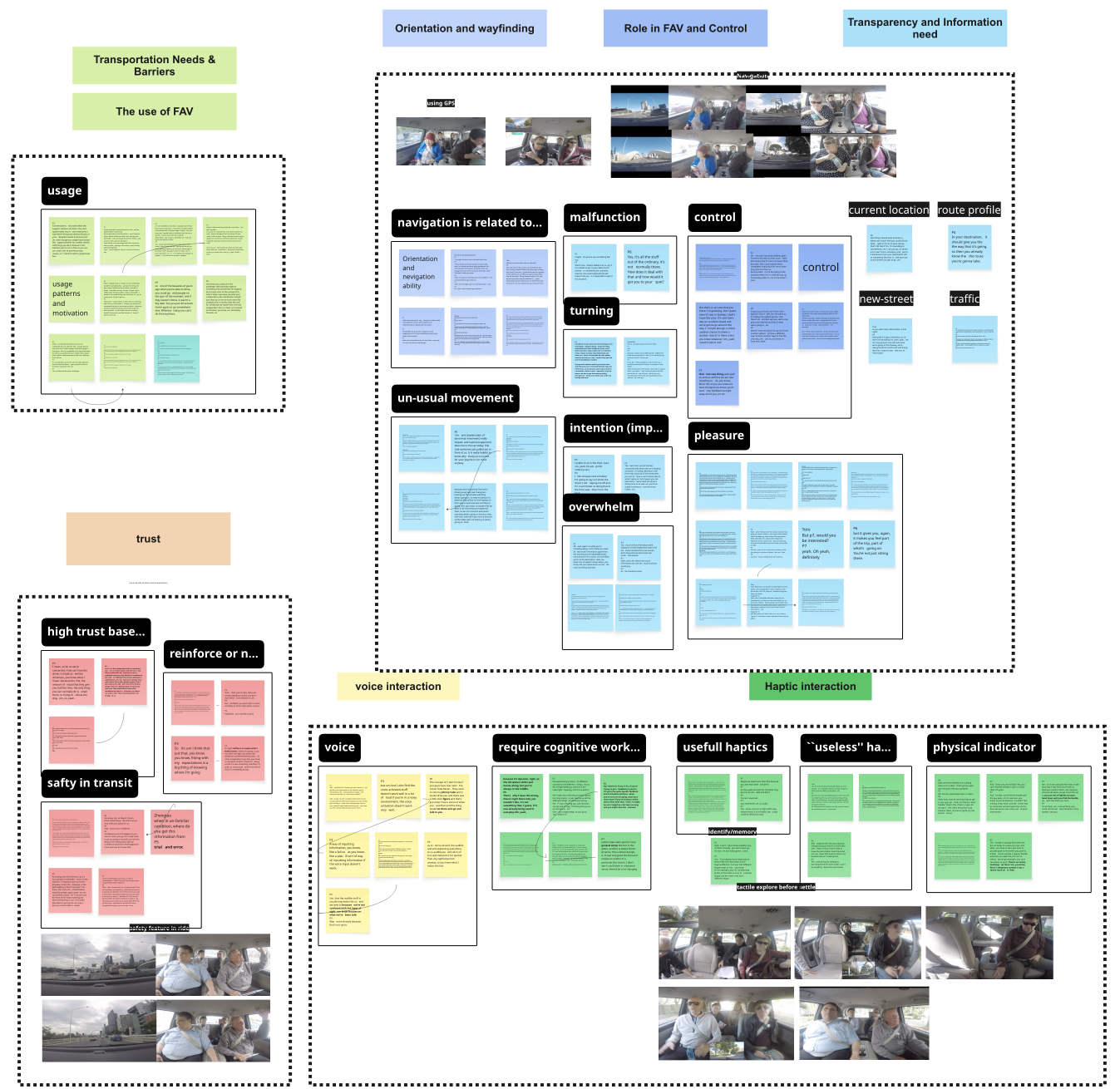}
    \caption{Affinity diagramming in Miro.}
    \Description{A screenshot of an affinity diagram in Miro, summarizing identified patterns with codes and coded segments. The diagram is organized into color-coded categories with sticky notes and example photos from the study. Many color notes are clustered into groups; connectors indicate relationships between codes. }
    \label{fig:Affinity diagramming}
\end{figure}

\begin{figure}
    \centering
    \includegraphics[width=0.85\linewidth]{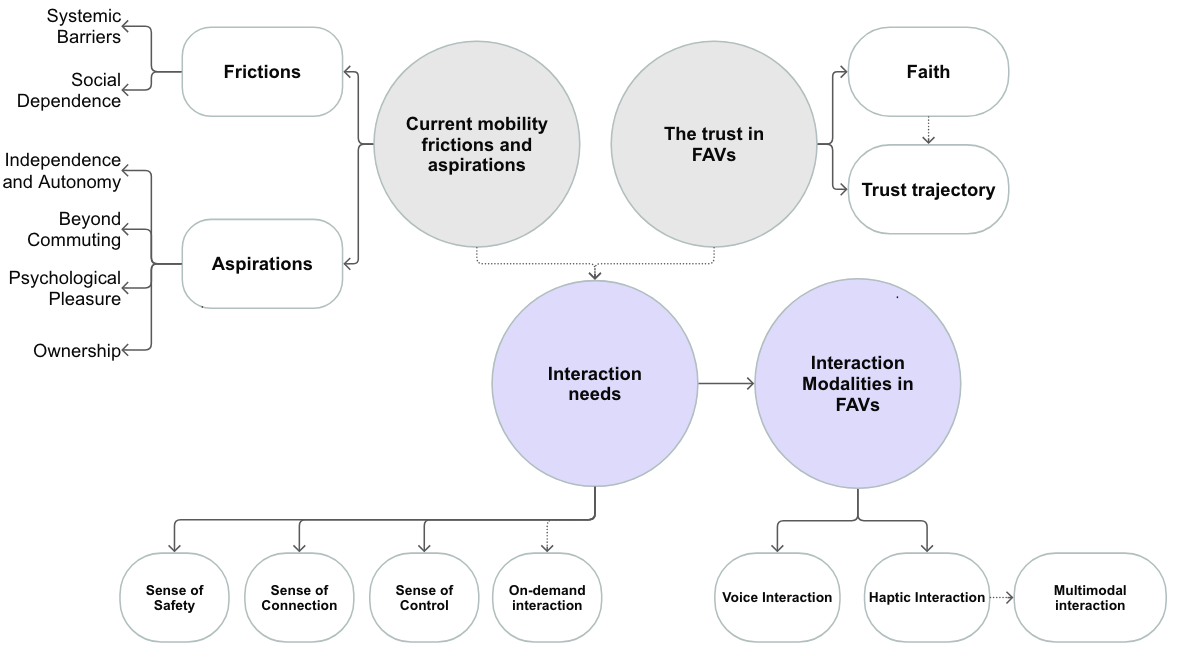}
    \caption{The generated themes and sub-themes. }
    \Description{Conceptual framework diagram showing relationships among themes of mobility, trust, and interaction needs in fully automated vehicles (FAVs). On the left, two branches feed into Current mobility frictions and aspirations (grey circle). Frictions include systemic barriers and social dependence. Aspirations include independence and autonomy, beyond commuting, psychological pleasure, and ownership. On the right, The trust in FAVs (grey circle) is linked to Faith and a Trust trajectory. Both grey circles connect downward to Interaction needs (purple circle), which expands into: sense of safety, sense of connection, sense of control, and on-demand interaction. Interaction needs lead to Interaction modalities in FAVs (purple circle), which expands into: voice interaction, haptic interaction, and multimodal interaction.}
    \label{fig: thematic analysis}
\end{figure}

We applied a thematic analysis following the approach outlined in previous research \cite{10.1145/3706598.3714205}, focusing on audio transcripts, with video as a complementary source. Three coders conducted collaborative open coding and affinity diagramming. They all have expertise in AV research, each with different disciplinary foci (product design, computer science, and human factors). They independently coded a 40-minute subset, and then discussed the initial codes to establish a shared common understanding. This resulted in 67 preliminary codes, which one of the coders (first author) applied to all remaining transcripts, while the other two coders applied the codes to half of the remaining transcripts each. All three coders allowed new codes and patterns to evolve. Video segments were selectively coded by the first author, guided by timestamps noted during transcript coding. In total, 591 coded segments were collected. Using affinity diagramming in Miro, codes were clustered into higher-order categories, supported by illustrative quotes and video stills, as shown in Fig\ref{fig:Affinity diagramming}. Through iterative discussion among all coders and authors, these were consolidated into a final set of four overarching themes. The themes and subthemes generated from participants' quotes and video are structured as follows, shown in Fig\ref{fig: thematic analysis}. 

\section{Findings}
Below, we describe our findings of BLV people's needs and interaction preferences in four themes: 1) Current mobility frictions and aspirations, 2) Trust in FAVs, 3) Situated interaction needs in FAVs, and 4) Interaction modalities in FAVs.  

\subsection{Current mobility frictions and aspirations}
BLV participants described their current mobility as constrained by both social dependence and systemic barriers. In our participant group, many relied heavily on their partners, children, or support workers to drive them places, which created feelings of inconveniencing others. As P13 explained, \textit{"I have to rely on someone who can see, right?...I feel like, you know, too much."} Others noted that public transport often failed to accommodate their flexible schedules, citing fixed routes, limited service times, and the high cost in both time and money. These frictions underscored participants' strong aspiration for greater independence and autonomy. 

FAVs were widely imagined as a way to restore that independence and broaden life opportunities. 
Several participants linked access to FAVs to improved quality of life through family connections and employment. For example, P7 envisioned, \textit{"You'd have the added advantage of seeing the grandchildren more frequently instead of once every three months,"} while P5 argued that \textit{"the chance of gaining employment would be 10,000\% increased just by having that availability of vehicle."}

\textit{Aspirations beyond commute.} In discussing how they might use FAVs, participants identified a wide range of scenarios beyond routine commuting. Daily tasks such as shopping or attending appointments were commonly mentioned, alongside longer trips to rural areas where public transport was limited, as P3 envisioned: \textit{"It'd be a great thing, particularly when you're going out of the city where the public transport routes don't go, for across-town sort of stuff."} Some anticipated shared use with family or support workers, emphasizing the social benefits of reducing the "onus" on others: \textit{"I'd like to take my partner out so she didn't have to drive. That'd be nice to do, to not have the onus on someone else to get you. I can pick you up. That would be cool."}(P6)
Others highlighted the hedonic dimension of `driving' itself, whether reclaiming the freedom they once enjoyed or imagining it for the first time. As P7 reflected, \textit{"The freedom of driving your car is fantastic. That’s so home,"} while P1, who had never driven, admitted, \textit{"I'm always dreaming about driving everywhere… I'd wanna sit in the driver's seat, and I'd wanna feel it, go righto, this is what it's gonna feel like just to drive in one of these self-driving cars."}

Ownership emerged as a critical aspiration. Private FAVs were viewed as the most desirable path to independence, even as participants acknowledged cost barriers. P1 speculated, \textit{"Unless they start issuing licences on how well we handle them,"} while P7 challenged policy restrictions: \textit{"If they're gonna have automated cars like Uber systems without a driver, then why can't a blind person have one and have a car?"} At the same time, participants recognized the affordability of ride-hailing models: P11 noted, \textit{"If it's a taxi, I certainly would try it, and if it was comparable to the half fare that we get."}

Taken together, participants envisioned FAVs as more than a transport service. They were described as tools for independence, platforms for social participation, and even spaces for psychological pleasure and self-realization. Independence was consistently framed both pragmatically (access to work, services, and everyday errands) and hedonically (connection with family, relief of family burden, and the embodied joy of "driving again").

\subsection{The trust in FAVs}

A distinctive theme in participants’ accounts was the way they positioned trust in FAVs. BLV participants described trust as a "faith" already woven into their daily mobility. Several explained that, depending on others, whether taxi drivers, support workers, or even strangers, meant “handing over” trust was a familiar, even habitual. As P6 put it:
\begin{quote}
\textit{``We hand over trust to somebody else every day. Whether it's walking on the road, or walking in a national park... You hand over absolute trust, with your life, to other people anyway. That's just what we do. And that's one of the hardest things, I think, for sighted people, to give over that trust. If you blindfolded them and put them in a driverless car, they're not used to that. That's something that's not foreign to us.''}
\end{quote}
For some, this orientation made the idea of a driverless vehicle feel less foreign. P5 reflected on the similarity between ride-hailing and automation: \textit{"As far as we're concerned, if we can’t see the driver, it might as well be driverless… half the time, the only thing you can normally do is smell them or try to refuse the dog."} In this sense, transferring trust from a human driver to an automated system felt like a continuation of their everyday practice rather than a radical shift. 

Participants also used the term "blind faith" themselves, reclaiming it as both a description of their current mobility experience and a resource for engaging with FAVs.
P10 described riding with unfamiliar drivers as \textit{"blind faith,"} while P8 added, \textit{"It is blind faith for sure. I think that’s the thing — when you can’t see you do have to look blind faith."} Importantly, these statements were not naïve: they revealed a pragmatic resilience shaped by lived reliance on others.

At the same time, participants emphasized that this trust was not limitless. It was maintained through experience, and small breaches could quickly undermine confidence. P6 noted, "\textit{You have little incidents, but I mean everybody does...you just hand over that all the time.}" Yet others admitted that repeated negative experiences could "knock your confidence," as P5 explained: "\textit{Once it happens to you once or twice, you go, do I really need to go out today or should I just sit here because I’m still trying to get my confidence back.}"

This dynamic points to a distinctive trust trajectory for BLV passengers. Whereas sighted users often require gradual exposure to automation before adopting it, BLV participants described beginning from a higher baseline of "faith", sustained by necessity and resilience. However, because in FAVs the accountable driver figure is absent, any unexplained maneuver or "bad behavior" risks being generalized to the FAV as a whole rather than to an individual driver. This makes timely explanations and consistent behavior critical to sustaining BLV users’ confidence. As P8 noted, "\textit{I think a driverless car, you’d have to build up a confidence and trust in it… you’d want to be aware, just so you know, because you’re throwing all your trust in it.}" In sum, participants described trust not as passive acceptance but as an active ability developed through lived experience. Their accounts suggest that FAV design must recognize this resilience while also supporting trust calibration. 

\subsection{Situated interaction needs in FAVs}\label{Interaction needs in FAVs}
Participants articulated a set of layered needs that shaped how they imagined interacting with FAVs, directly addressing RQ1. These needs extended beyond pragmatic safety to encompass social connection, psychological comfort, and a desire to remain engaged as active participants in travel. We organize these needs into three core dimensions: safety, connection, and control, with an additional emphasis on on-demand interaction to avoid information overload or feelings of being mocked. 

\subsubsection{Sense of Safety}
Safety was the most fundamental layer of need, particularly during transition phases such as locating, boarding, and exiting the vehicle. These moments were described as especially vulnerable without a sighted companion. As P8 explained, “\textit{When you’re in the car, that’s your safe bubble}," and P7 added, "\textit{But when you’re getting out...so of course the self-driving car would not know there’s a pole gonna block the door. So it means we’d have to take extra care.}” 

Participants stressed the importance of reliable cues for approaching the vehicle, locating door handles, and confirming whether doors were locked or could be opened in emergencies. Identifying a stationary, silent car was seen as particularly challenging. As P11 noted, “\textit{They are silent once the engines stop...I’ll have to watch, find that video to see how she found the Waymo.}” Some expressed frustration with existing solutions. P10 critiqued the honk-based identification used in some robotaxis: “\textit{By the second honk, I’ve sort of focused where the noise comes from, and then there’s nothing. I would need constant noise or something.}”

Design features considered “convenient” for sighted users could create hazards for BLV passengers. P5 warned, “\textit{Things that would be really dangerous in a driverless taxi, like automatically opening a door...for me, it’d probably open it into me or I’d run into it expecting it to be closed.}” Similarly, P14 raised concerns about accidents and emergency egress: “\textit{If you had an accident and the doors were locked...you couldn’t light from the vehicle [to open the door].}” These accounts underscored the need for FAVs to provide dependable and accessible mechanisms for both everyday boarding and emergency exit. Video review further showed how participants tactually explored doors during the study, actively using touch to verify safety features. This behavior highlights the embodied strategies BLV passengers rely on.

Safety also extended to disembarking and navigation after drop-off. P5 highlighted, “\textit{Is it going to pull up on the driveway, across the driveway, at the path leading to the front door? That sort of information would be greatly appreciated.}” In-ride safety was shaped less by concerns about mechanical failure than by the vehicle’s driving behavior. Smoothness, predictability, and clear cues of intention (e.g., indicator clicks) reassured participants that the vehicle was acting responsibly.

\subsubsection{Sense of Connection}
\begin{figure}
    \centering
    \includegraphics[width=0.75\linewidth]{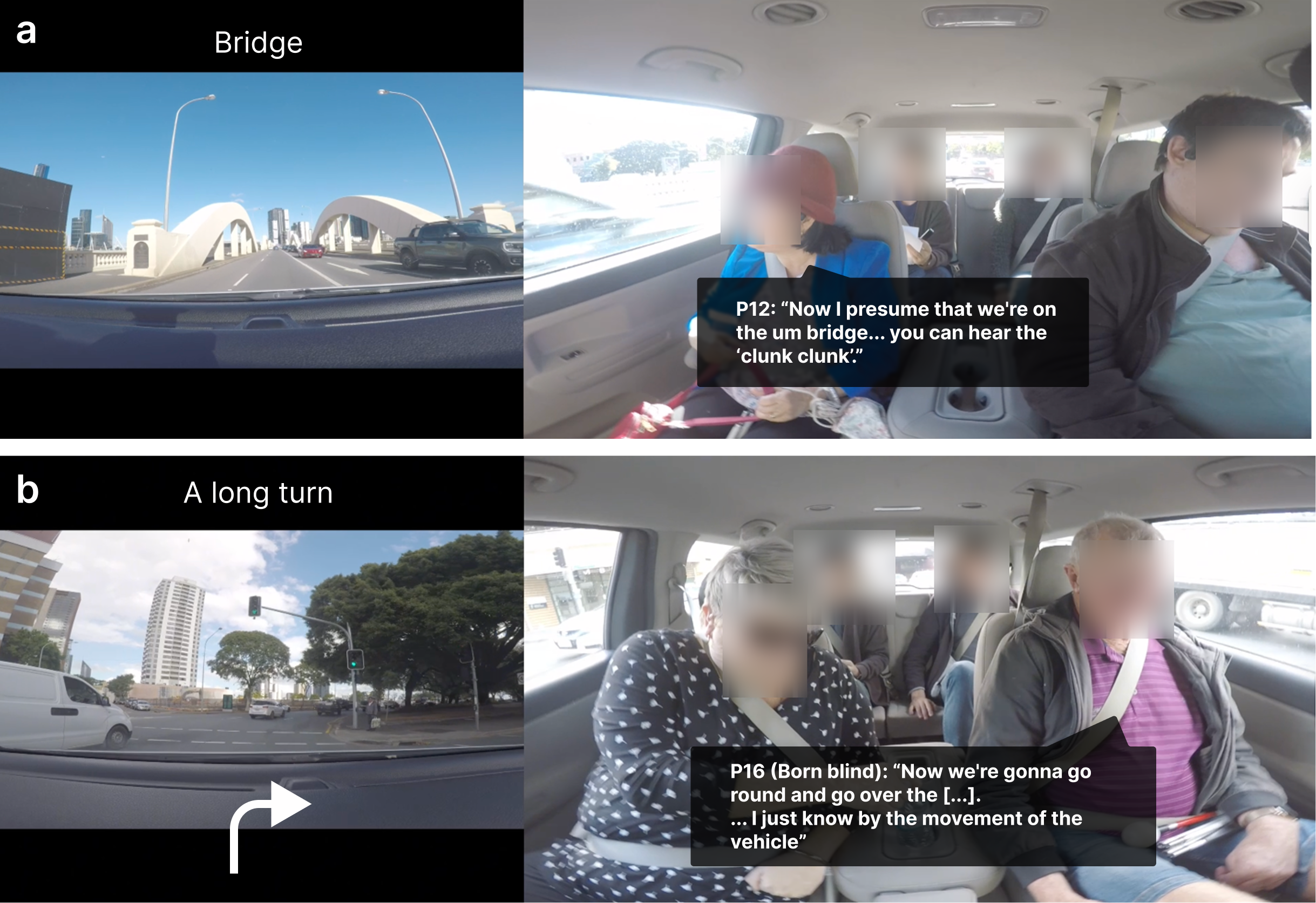}
    \caption{Subconscious connection to familiar routes. Users' navigation by recognizing characteristic vehicle movements and sounds.}
    \Description{Two sets of paired photos showing participants in the on-road study interpreting environmental cues. (a) Bridge crossing: Left image shows the car’s forward view of an arched bridge with vehicles ahead. Right image shows participants inside the vehicle; one passenger in a red hat is speaking. Overlaid quote from P12: “Now I presume that we're on the um bridge… you can hear the ‘clunk clunk.’ (b) Long turn: Left image shows the vehicle approaching a wide right turn at a city intersection with traffic lights and tall buildings. Right image shows participants seated inside the car, one passenger in the front seat speaking. Overlaid quote from P16 (Born blind): “Now we’re gonna go round and go over the […]. I just know by the movement of the vehicle.”
}
    \label{fig:Users' navigation}
\end{figure}

Participants wanted to remain connected both to the journey and to the surrounding environment. A subconscious connection to familiar routes and environments was evident among participants; as shown in Fig.~\ref{fig:Users' navigation}, they oriented themselves through characteristic vehicle movements and sounds.
Strategic information such as route profile, estimated duration, and key street names was valued, as it supported orientation and reduced feelings of isolation. P3 noted, "If it was a driverless car, I'd want it to be telling me...we are now going to turn left onto here, we're going on the freeway, we're taking the Brian Smith's exit." 

Connection was especially valued when unexpected events occurred.
Participants desired explanations, either actively asked or passively received, to maintain trust and orientation. As P2 explained, “\textit{If we suddenly hard brake, it would be good to ask the interface, ‘what just happened?’}” Similarly, P6 recalled reassurance from a human companion: “\textit{She said someone just pulled out in front of us. It is really helpful to know why that just occurred for your psyche.}”

Views on environmental information varied. Some appreciated incidental or descriptive details, while others saw them as unnecessary or overwhelming. P6 expressed enthusiasm: \begin{quote}
    “\textit{My partner will just go, ‘They’re pulling the building down over there.’ It helps make your trip something, as opposed to just sitting there. It just helps to be part of inclusion in where your surroundings.}”
\end{quote} Others preferred such input to be optional, as P7 put it: “\textit{People wouldn’t want that running all the time. You just say tourist mode, and it starts to describe things going on.}”

Beyond explicit system outputs, participants also noted that incidental sensory cues (wind, smells, and ambient sounds) enhanced comfort and fostered connection to the ride. While subtle, these accounts suggest an opportunity for FAV design to intentionally preserve or amplify such sensory richness, supporting not only orientation but also ride enjoyment. 

\subsubsection{Sense of Control}
Control was framed not in terms of manual driving, but in being included in higher-level strategic decisions. Participants wanted to be “in the loop” for detours, delays, or rerouting. P6 explained, “\textit{If it tells me there’s an accident up here, would I have an option to say: ‘Is there another way?’ Even if I didn't know the area, I might also go: `is there another choice?'}” Participants emphasized that control should be a “\textit{two-way thing},” appreciating continuous interaction rather than a one-time selection made at the start of the trip.

For some, control also had an experiential dimension: sitting in the driver’s seat or feeling the car’s movement symbolized regaining the lost identity of being a driver. P2 remarked, “\textit{When I’m in a self-driving car, I’d actually prefer to be in the driver’s seat.}” Similarly, P1 acknowledged this as more of an experiential preference than a practical requirement, describing how tactile feedback could simulate driving: 
\begin{quote}
    "\textit{I always dream of riding motorbikes too...they make Mini Guides\footnote{Mini Guide is an electronic mobility aid and uses ultrasonic echo location to detect objects. The aid vibrates to indicate the distance to objects, the faster the vibration rate the nearer the object.} The same principle the mini guides made on the laser, so why couldn't they put that into the handlebars of a bike? Potentially if you could do that in the steering wheel... Yes, you'd still want to put that computer assistance, but part of me always dreams of having some control.}" (P1)
\end{quote}
Together, these accounts show that control for BLV passengers is less about direct vehicle operation than about participation and presence: being consulted in strategic decisions and, at times, reclaiming the embodied experience of driving.

\subsubsection{On-demand interaction}
Participants consistently emphasized that information should be available when needed, rather than constantly broadcast. Over-communication was often described as stressful or even patronizing. As P10 put it, “\textit{We want the information on demand. We don’t want it constantly.}” Similarly, P8 explained, “\textit{That [constant information] causes stress and anxiety, to be honest.}” 

Irrelevant or non-actionable details were particularly frustrating.
As P2 asked, “\textit{For what purpose is it telling me, unless I have to do something about it?}” P3 critiqued overly granular environmental updates: “\textit{Like telling you 6 o’clock there is a bicycle...I’m not interested. It might have been interesting, but I don't want to be, it's almost like it's mocking me.}” P5 echoed this sentiment: \begin{quote}
   "\textit{I guess that's just one of the things that you lose when you lose your sight. because it could describe it to you, but you can't see it, it almost like it's mocking. Some people may like it, well, that might give you a marker as to where you are in relation to your journey or something, but if you can't see it, it's like, yeah, well, thanks very much.}" 
\end{quote} Such comments reveal that information which cannot be acted upon is experienced less as helpful orientation and more as a reminder of exclusion. 

These accounts underscore that FAV interfaces must support adaptive, on-demand information delivery: providing actionable cues proactively, while leaving contextual or hedonic information accessible at the passenger’s request. This ensures not only efficiency but also dignity, by avoiding overload and resisting designs that feel condescending.



\subsection{Interaction modalities in FAVs} \label{subsection:Interaction preferences in FAVs}
Having outlined the layered interaction needs of BLV passengers during FAV travel, we now turn to address RQ2 by examining how these needs intersect with preferred interaction modalities, particularly haptics.
\subsubsection{Voice interaction}
Voice interaction was the most frequently preferred modality, described as both simple and effective. Participants valued it for offering direct access to information and immediate action without the need for touch-based interfaces. As P9 explained, “\textit{Voice for a driverless car to tell us where we are and tell us information about our surroundings.}” For many, this preference reflected familiarity and efficiency in everyday life, where voice already supported communication with digital assistants, screen readers, and interpreting services.

\textit{Voice as a competence.} Participants underscored that reliance on auditory input was not a constraint but a cultivated competence. P5 recalled high-speed guided driving as evidence of distinctive expertise: "\textit{The race tracking was all done guided by hearing. Instructors said that our ability to understand and interpret and do what we were told, they'd never come across sighted people who were able to do it...Across with blind people, their audible interpretation skills are in the top 1\% of the population.}" Such accounts framed voice interpretation as a marker of pride and competence, suggesting that voice interaction aligns with BLV users' embodied capabilities. 

\textit{Voice agent.} Participants also discussed qualities that shaped the usability of voice agents. They stressed the importance of continuous, two-way interaction rather than one-off commands (P6), and called for smart assistants trained with driving-specific language, as P9 said, "\textit{with keywords built into it for a car, not for a phone.}" Voice quality mattered for engagement during interaction: P14 compared poorly narrated audiobooks with the need for a "\textit{decent voice}" assistant. While some imagined more embodied presences, such as talking figures ("\textit{the Johnny Cab robot\footnote{The automated taxi driver in the 1990 film Total Recall.} would sit there and go and talk to you.}" (P5)) inside the vehicle to offer social presence, these were framed as cultural references rather than concrete requirements.



\textit{Collaboration with human interpretation.} Beyond voice agent, participants described how existing human interpretation services (e.g, \href{https://www.bemyeyes.com/}{Be MY Eyes}, \href{https://aira.io/}{Aira}) could be integrated into FAVs to provide both functional support and companionship. P6 described this integration as both functional and social: 
\begin{quote}
    "\textit{Like if I could call Aira on my phone, but it uses the camera of the car, Um, and they can talk to me about where we, what we're doing, or even if I just wanted to have a chat.}" 
\end{quote} Here, participants described trusted external resources as a way to go beyond built-in HMIs, supporting safety and orientation while also offering social connections that made the ride more engaging.

Despite strong preferences, participants acknowledged that voice could not be universal. As P7 noted, “\textit{maybe we need more than that (voice interaction) because you also have deaf and blind,}” and P3 added, “\textit{if you're in a noisy environment, the voice activation doesn't work very well}.” These concerns underscore that while voice aligns with BLV competence, its reliability is context-dependent and must be complemented by alternative modalities.

\subsubsection{Haptic interaction}
Haptic interaction drew diverse opinions, with its usability perceived as highly context-dependent. Participants recognized value in tactile cues when these were directly interpretable and tied to embodied action, but raised concerns about designs that lacked tactile affordances or introduced abstract signals with little practical meaning. 

\begin{figure}
    \centering
    \includegraphics[width=0.85\linewidth]{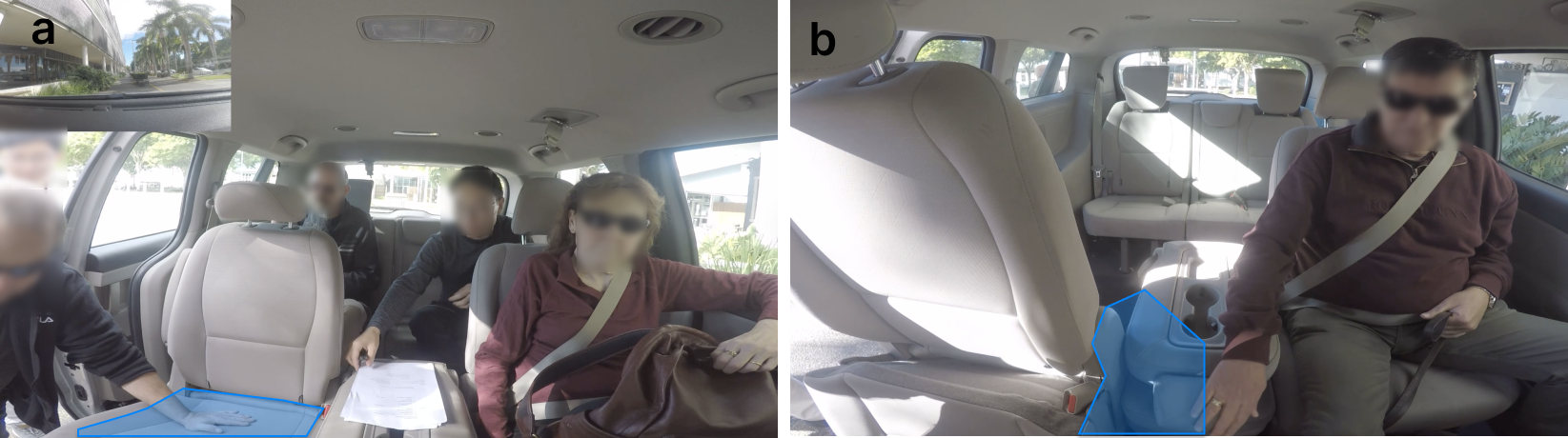}
    \caption{A sample of participants’ tactile space scanning before settling in. (a) A participant swiping his hand across an empty seat before sitting on it. (b) A participant exploring the space on the right side by touching adjacent empty seats after sitting. }
    \Description{Two photographs showing blind and low-vision participants seated inside a people mover during the on-road study. (a) Participants and researchers are seated in the 2nd and third row seats; one participant was exploring the vehicle interior by touch before entering. An inset view in the top left shows the outside street and building. A blue-highlighted area on the empty seat marks a tactile exploration zone.(b) A single participant sits in the 2nd passenger seat, touching the blue-highlighted area on the right side of his own seat. The back seats are visible in the background, illuminated by sunlight through the windows.}
    \label{fig:tactile space scanning}
\end{figure}

\textit{Embodied confirmation.} Participants frequently relied on touch to orient themselves, both in routine exploration and during the study rides. Video review showed how participants scanned doors, handles, and seats (own, next, front) to confirm their relative location and ensure safety before settling in, shown in Fig\ref{fig:tactile space scanning}. This active tactile exploration underscored that haptics was not an additional input channel but an existing, embodied strategy for interacting with vehicles. P7 contrasted this with designs that obstructed tactile access, citing Tesla’s flush handles as an exclusionary feature:
\begin{quote}
"\textit{What we do when we get out of the car is you just sort of lightly run your hand down until you find the handle, uh, with the Tesla, you can't, you could not feel uh any indication of the handle.}" (exit)
"\textit{Unless you've got eyesight, you can't see where to push. There are no tactile markings on them, like if a telephone number 5 has a tactile mark on it.}” (entry)
\end{quote}
This example highlighted the risks of removing tactile reference points, where design decisions made for aesthetics or aerodynamics inadvertently undermined safety and usability for BLV passengers. 

\textit{Navigation and orientation.} 
Some participants suggested more proactive haptic feedback, such as shape-changing for a tactile display, for supporting navigation and orientation. P1 (who is born blind and with Braille literacy) described the potential of tactile maps: "\textit{It might be worth looking into that Braille notes and things, and seeing whether that technology can somehow be incorporated into it. Then you might be able to get a diagram of the map on the screen, um, that can come up and down...}" These tactile graphical renderings were valued primarily for providing precise, localized confirmation (e.g., parking spots) rather than for supporting continuous in-ride navigation. P2 explained:  
\begin{quote}
    "\textit{When you're driving along, I don't think it [tactile map]'s gonna be a lot of use. But certainly when you're coming to (exit), you're trying to pull into a parking space, and you can actually feel, oh, there's 3 houses along. So, you know, 'OK, let's head for that one.'}"
\end{quote}
However, there are more skeptical opinions towards this idea of using tactile for representing roads and surroundings, considering the added hardware complexity, interpretation difficulty, and adding extra cognitive load without clear added value. P3 reflected: "\textit{There's too much interpretation, particularly if you're in, if it's an unfamiliar area. Um, I don't think it's much use.}" Similarly, P16 remarked, "\textit{I can feel that these roads are parallel and running across me, but what are their names? ...if you're in an area where we're not familiar, it's just like a puzzle}." Such comments suggested that speculative uses like tactile maps might offer value for specific, localized tasks, but were unlikely to support continuous navigation. 
\begin{figure}
    \centering
    \includegraphics[width=0.75\linewidth]{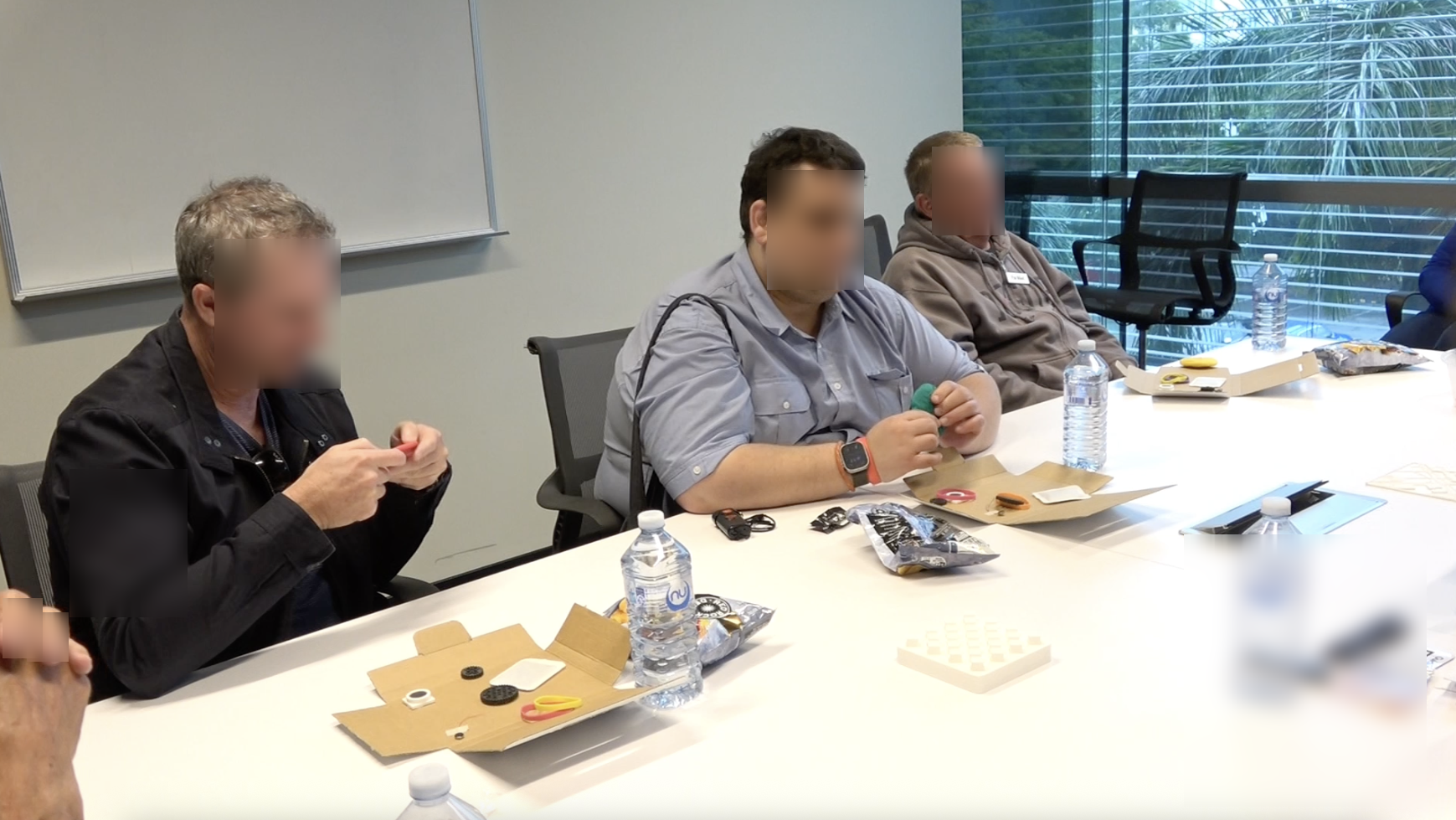}
    \caption{In-lab focus group with haptic probes. Participants exploring Play-Doh as a deformable, shape-changing interface.}
    \Description{Three participants seated around a white table in a meeting room during an in-lab study. Each participant has a cardboard tray containing small haptic materials, such as rubber bands, plastic gears, and clay. Two men in the foreground are actively handling modeling clay and other tactile objects, while another participant sits further back. Bottles of water and snack bags are provided on the table.}
    \label{fig:In-lab focus group with haptic probes}
\end{figure}

\textit{Critiques on cognitive load.} Participants stressed that haptic cues need to be simple to interpret and low in memory load. Otherwise, they risked confusion, misinterpretation, or emotional frustration. P3 captured this tension: "\textit{If you're having to think about what does that mean, it might be...too long and you might miss the moment.}" Similarly, P4 described the problem of memorizing patterns: “\textit{It gets too confusing and you spend too long trying to just memorize all the different things. Give you a beep beep, so you go to say: what is it?}” If too many haptic signals felt similar, participants feared misinterpretation.  

\textit{Critiques on actionability.} Several participants criticized non-actionable haptics, such as seat vibrations indicating nearby traffic. P3 noted: "\textit{Suddenly the seat vibration changes because there's a semi-trailer over there. I mean, I don't need to know the semi-trailers over there...The presuming the software is worrying about the semi-trailer. I can't do anything about it.}" P2 summarized: “\textit{Since we can't do anything about it...for what purpose is it telling me, unless I have to do something about it?}” These accounts underscored that “useless” haptics not only add cognitive workload but also generate negative emotions, making participants feel anxiety and helplessness. Participants stressed that effective haptic signals should be quick to identify and be clearly tied to actionable information. 

\textit{Tactile literacy and familiarity.} Assumptions about tactile literacy also limited the feasibility of some haptic proposals. While P1 expressed interest in Braille-based maps, P3, a teacher, countered: “\textit{Braille almost superfluous...it’s not very practical...most kids once they finish school...they rely so much more [on] voice.}” While tactile maps were considered useful for grasping general sense (e.g., "\textit{like this is the globe, and this is where North America}" (P3)), they were less practical for complex or dynamic representations, such as traffic.
Especially when the haptic is in a dynamic sense, these "novelties" would be foreign to BLV users, as it is not common in their daily Assistive technology use. P5 explained, "\textit{It's not something that was already being used in everyday life. You [sighted users]'d already have an idea of how it works, you could educate blind people on it, but for what purpose really?}" 

Other novel haptic modalities, such as thermal or force cues, were also less familiar. P3 dismissed thermal cues: “\textit{`Something's on fire!' you know...that would be non-specific feedback.}” P16 offered a more positive spin, suggesting heat might indicate time of day: “\textit{That also helps, you know what time of day it is,}” recalling how he oriented himself by the sun’s warmth. Despite such speculation, thermal feedback was not broadly embraced. As P8 summarized, “\textit{The more familiar environments, the better it is for us...the more we feel more safer.}”

\textit{Multimodal interaction.} When discussing multimodal interaction, participants emphasized the importance of evaluating whether adding a haptic channel has increased value. P8 questioned, "\textit{I questioned is it(haptic) add value? And for me, I don't think it would...when you're not decision making in you, when you can't see, you're already processing a lot...you don't need another level of complexity.}" 
Others, however, acknowledged the importance of having multiple options for accessibility. P9 noted, "\textit{having different ways of interaction, so you'd have voice, you would have buttons, or you'd have some form of tactile, uh, integration...}" Overall, they think no matter voice or other modality, they highlight two principles: consistency and standardization to make use of or enrich familiarity. 

\paragraph{In summary} Participants positioned haptics as valuable when they aligned with embodied strategies already in use, such as touching doors and seats to confirm safety, or when they offered localized confirmation like parking orientation. However, speculative extensions such as tactile maps raised concerns about literacy, hardware complexity, and cognitive load. Abstract or non-actionable haptics were often experienced as confusing or even frustrating, while unfamiliar modalities such as thermal feedback held little practical appeal. 

\section{Discussion}
Our study provides evidence of BLV users' attitudes and interaction needs in FAVs, specifically how their interactions reveal safety, trust, and autonomy values of BLV users in FAV use. We discuss these findings with previous literature, and ground a dignity perspective behind accessibility.  

\subsection{BLV users' faith}
%
Our data suggest a starting point for many BLV riders: a baseline of \textit{"faith"}, an important stance developed both from the promised mobility benefits of FAVs and lived experience with habitual uncertainty in mobility. As P6 noted, “we hand over trust… every day.” Rather than naïve or “blind faith” in the participants' quotes, they described this "faith" as a situated strategy shaped by the necessity of mobility, consistently outweighing potential frictions or even risks. 

Similar dynamics are reported in BLV ride-sharing studies: when negative driving happened, BLV riders' experience-based trust is eroded, yet they often continue using the service because independence and social participation outweigh isolated failures; critically, distrust is attributed to the driver, not the entire system \cite{kameswaran_we_2018, brewer_understanding_2019}. We questioned whether this "faith" that participants mentioned is resilient, since the driver is removed. In FAVs, accountability and reassurance move from a human in the driver's seat to a whole automated system (including in- and exteriors, automated driving behaviors, HMI, and back-up safety assistance). Consequently, a single adverse experience (hesitant maneuvers, unexplained hard braking, confusing pickup) risks being generalized to the system as a whole rather than to an individual driver. 

Many participants perceived faith as helpful for both initial adoption and maintaining psychological comfort during FAV travel, although these assumptions warrant further examination in studies with real FAVs. Brinkley et al.'s in-lab focus group likewise observed that some blind participants were willing to “take the risk,” motivated by optimism that FAVs could potentially be life-changing technology for them personally \cite{brinkley_opinions_2017}. Our on-road study grounds this optimism in real driving conditions, but also reveals the heightened need for timely explanation and predictable recovery to preserve riders' confidence. In Muir and Moray's early automation trust model, they distinguish an affective faith component that can seed early acceptance but must be supported by predictability and dependability to mature into sustained trust \cite{muir1996trust}. For BLV users, participants' recurring “what just happened?” requests point to a need for situation awareness via concise, attributional messages. 

We see the opportunity to treat BLV users’ "faith" in FAVs as a grown ability. Designing for this ability requires interactions that provide predictability cues (consistent interaction flows, pre-education or instruction on FAVs), dependability cues (action confirmations and opportunities for control), and trust recovery mechanisms (explanations after anomalies). Additionally, the interaction given should respect BLV passengers’ "faith": not every uncertainty must be eliminated, but support calibration by making system intentions and next actions legible at the moment they matter.



\subsection{Layered interaction needs}
Participants described their interaction needs in FAVs as layered, spanning safety, connection, and control. This sequential structure was reflected in the way concerns emerged during discussions: all participants immediately raised issues of safety and risk, later emphasized connection needs while riding, and some of them articulated control needs when prompted by researchers.

Safety confirmations in FAVs such as locked doors, fastened seatbelts, and reliable egress largely mirror prior findings that highlight safety as a baseline requirement for BLV mobility \cite{brinkley_exploring_2020,fink_fully_2021,fink_give_2023}. Similarly, connection has been discussed in work on interfaces that expand situation and environmental awareness \cite{fink_expanded_2023,meinhardt_hey_2024}. Our participants added that incidental sensory cues (e.g., airflow, smells, and ambient sounds) also contributed to comfort, grounding this value of travel pleasure. Control, widely recognized as a desired power of BLV riders \cite{brewer_understanding_2018}, was here expressed as the need to be offered concrete options: querying detours, confirming or rejecting route changes, and staying “in the loop.”

However, participants consistently rejected the idea that accessibility means “more information.” Instead, they called for concise, on-demand, and actionable cues.  While under-communication hindered the development of trust and harmed BLV passengers' "faith", over-communication was commonly seen as too much cognitive workload, which is also one main criticism of interaction design in AVs \cite{cognitiveload_avs_2014}. 

Participants highlighted that poorly calibrated delivery could lead not only to overwhelm but to negative emotions. Some perceived excessive updates as “mocking,” while others noted frustration when information was irrelevant to action: “for what purpose is it telling me, unless I have to do something about it”. These accounts reflect a need for the respectful design philosophy that attends to the emotional dimension of interaction \cite{Oro25_dignity}: information mismatched with agency or aesthetics does not merely add noise, it undermines dignity.

\subsection{Designing for dignity: beyond access}
When considering the needs in FAVs, participants asked for more than access: they emphasized being informed equally as sighted users, retaining in‑loop choices, enjoying the cabin or trip as a hedonic and social fulfillment, and being respected for their abilities while interacting. Drawing from the principles of dignity-oriented design by Colhlan et al. \cite{Dignity_Coghlan_2021}(dignity, autonomy, and company), our findings suggest that interaction design in FAV for BLV users should extend beyond basic functionality to affirm their psychological well-being. Evolved from BLV users' perspective towards interaction needs, we explained the dignity dimensions in FAV accessibility: \textit{(1) Affirm autonomy, (2) Minimize intrusive load, (3) Respect competence, (4) Enable emotional and social fulfillment, and (5) Support healthy interdependence}.

(1) Interaction in FAVs should \textbf{affirm autonomy} by keeping BLV users “in the loop” at strategic decision points, such as route choices and timing approvals. This supports a stronger sense of independence with the “power of control,” \cite{brewer_understanding_2018} while also ensuring safety.

(2) Interaction in FAVs should \textbf{minimize intrusive cognitive load}. Lived experience with uncertainty is a defining aspect of BLV users’ technology use \cite{EverydayUncertainty_Tang_2025}, and participants in our study also show how this extends to their mobility. Rather than projecting the information expectations of sighted FAV users, interaction design should minimize intrusive cognitive load by engaging with BLV users’ own perspective on the “unknown” and what forms of uncertainty are acceptable. 
This entails providing layered, value-sensitive information with clear opt-in/quiet defaults, ensuring the flow of information is helpful without being overwhelming, and, more importantly, avoiding patronizing. 

(3) Interaction design in FAVs should begin by \textbf{acknowledging the BLV user's competence} by leveraging skills they already possess \cite{Reyes_Competency_2020}, such as natural language, or their tactile exploration protocols close to the body. This is not merely a convenience but a mode of fluency and control that places them on an equitable footing with sighted riders. 

(4) Interaction in FAVs should also \textbf{facilitate BLV users' emotional and social fulfillment}, transforming the vehicle from a simple mode of transport into a hedonic and social space. While participants repeatedly emphasized the hedonic appeal of the “pleasure of driving,” FAV interactions could draw on this sense of driving to enhance user experience, for example, through vibrotactile cues on a steering wheel \cite{sucu_haptic_2013} or a tangible vehicle representation in hand \cite{meinhardt_hey_2024}. Beyond interaction for emotional fulfillment, FAV systems should also support BLV users’ social fulfillment, such as by enabling services or interactions that allow them to “drive” others. 
This dignity dimension emphasizes private ownership compared to assumptions about autonomous ride-sharing or ride-hailing services \cite{fink_give_2023}. In such contexts, users can engage in conversation, listen to music, and even recapture the “pleasure of driving.”

(5) Interaction design in FAVs needs to \textbf{support healthy interdependence} in mobility context. Rather than focusing on a single, isolated BLV user, the design should acknowledge that traveling with companions or support workers is a common and valued practice for BLV users, and is related to the destination activities. By facilitating a better shared use and collective participation, the FAV can redefine the user's role from a passive passenger to an active collaborator in a communal journey. While Fink et al. \cite{fink_give_2023} examined BLV users’ attitudes toward “human (driver) or AI,” our findings highlight the opportunity of “human \& AI”: delegating driving operations to automation while still embracing human support services, such as integrating Aira with FAV sensors. This approach ensures the FAV enhances their lives without isolating them, while respecting the interdependence between BLV passengers and their trusted source and companions \cite{Interdependence_Bennett_2018}.

In summary, our study situates the design of dignity in FAVs by grounding BLV interaction as an adaptive information ecology, a perspective that is instrumental in fostering dignified accessible FAVs.

\subsection{Reconsidering haptics value through BLV users' perspective}
In our studies, participants consistently preferred "simple" haptics (e.g., tactile differences on doors, physical markers on knob or screens) over expressive haptics (e.g., dynamic tactile maps, directional nudges, thermal dashboards). The latter were perceived as "like a puzzle" or burdensome. Similar tension is also found in early-mentioned haptic HMIs \cite{Fink_Autonomous_2023_midair, meinhardt_hey_2024}, where +haptic interfaces produced no significant improvements over voice alone. Our qualitative data articulate the cognitive and psychological reasons for this: haptics that require new interpretations or don't align with a user's existing mental models feel alienating. 

It is also important to understand the language of haptic habits that BLV users rely on in mobility. Although participants expressed varied feedback on haptics in FAVs, our findings revealed shared tactile scanning protocols that they consistently used to orient themselves in the car. In this sense, haptic interaction in FAVs should support and extend these existing practices, thereby defining the design space from the BLV user’s perspective. 

This distinction aligns with a broader rethinking of haptic experience \cite{HXD}: rather than solely focusing on technical novelty, designers should frame the haptic experience as user-centered clarity. 
While designing haptic HMIs for BLV users in FAVs, our study critiques haptics on interpretability, actionability, and familiarity. For BLV riders, haptic HMIs' usability hinges on interpretability: they wanted cues that are easy to decode while the vehicle is moving and avoided cues that added working-memory load or required visual modality to interpret (such as the malfunctioning signals in most vehicles). 
As for actionability, helpful haptic messages implied an immediate next step (e.g., “reach here”, “confirm locked”, and “vehicle ready”). Design also needs to respect and enrich BLV users' familiarity, acknowledging that haptic literacy varies across BLV participants. While aligning haptic cues with established tactile repertoires, brief instructions or human-guided practice are also suggested to scaffold learning when novel haptic signals are introduced. This approach aligns with our respectful design philosophy, ensuring that technology builds on users’ autonomy and abilities without compromising dignity.

\section{Limitation}
Limitations exist within this study. Firstly, while immersed in an on-road focus group, participants remained aware of the presence of a human driver resulting in discussions that may not have fully extend to perceptions of the vehicle’s driving ability, which may differ in a truly driverless context. Secondly, all participants were recruited from a single city, where local cultural norms and transportation practices likely shaped their responses, and findings may not generalize to BLV communities in other regions. Finally, participants attended the study in person, indicating a certain level of independent navigation ability. And as all participants were not newly blind, possessing a developed self-navigation strategies, the perspectives of individuals with recent vision loss who may not yet have established such skills are underrepresented in this dataset.
\section{Conclusion}
We conducted a series of on-road focus groups with blind and low-vision participants in a simulated fully automated vehicle to examine how interaction needs and modality preferences in FAVs are shaped by context and user competence. Through thematic analysis of ride experiences and sensory interactions, we demonstrate that layered information delivery supports BLV users’ values and affirms their dignity. Building on these findings, we articulated \textit{a design-for-dignity lens across five aspects of autonomous mobility} and reconsidered the role of haptics in FAVs for BLV users. We argue that this lens reframes inclusive FAV design and informs the development of respectful human–machine interfaces that strengthen mobility independence for BLV users. 
\begin{acks}
Anomized during review. 
\end{acks}

\bibliographystyle{ACM-Reference-Format}
\bibliography{sample-base}

\appendix
\end{document}